# Is the Structural Relaxation of Glasses Controlled by Equilibrium Shear Viscosity?

Ricardo Felipe Lancelotti[1,2,*], Daniel Roberto Cassar[2], Marcelo Nalin[3], Oscar Peitl[2], Edgar Dutra Zanotto[2,*]

[1]*Federal University of São Carlos, Graduate Program in Materials Science and Engineering, 13565-905, São Carlos, SP, Brazil.*
[2]*Center for Research, Technology and Education in Vitreous Materials, Department of Materials Engineering, Federal University of São Carlos, 13565-905, São Carlos, SP, Brazil.*
[3]*Institute of Chemistry, São Paulo State University, UNESP, 14800-060, Araraquara, SP, Brazil.*

\* lancelotti.r@dema.ufscar.br, dedz@ufscar.br

## Abstract

Knowledge of relaxation processes is fundamental in glass science and technology because relaxation is intrinsically related to vitrification, tempering as well as to annealing and several applications of glasses. However, there are conflicting reports—summarized here for different glasses—on whether the structural relaxation time of glass can be calculated using the Maxwell equation, which relates relaxation time with shear viscosity and shear modulus. Hence, this study aimed to verify whether these two relaxation times are comparable. The structural relaxation kinetics of a lead metasilicate glass were studied by measuring the refractive index variation over time at temperatures between 5 and 25 K below the fictive temperature, which was initially set 5 K below the glass transition temperature. Equilibrium shear viscosity was measured above and below the glass transition range, expanding the current knowledge by one order of magnitude. The Kohlrausch equation described very well the experimental structural relaxation kinetics throughout the investigated temperature range and the Kohlrausch exponent increased with temperature, in agreement with studies on other glasses. The experimental average structural relaxation times were much longer than the values computed from isostructural viscosity, as expected. Still, they were less than one order of magnitude higher than the average relaxation time computed through the Maxwell equation, which relies on equilibrium shear viscosity. Thus, these results demonstrate that the structural relaxation process is *not* controlled by isostructural viscosity, and that equilibrium shear viscosity only provides a lower boundary for structural relaxation kinetics.

# 1   Introduction

Any liquid that survives a cooling experiment below its liquidus temperature without crystallizing is denominated supercooled liquid (SCL). If cooled further, its structure eventually freezes at the glass transition temperature ($T_g$) as the thermal energy available becomes insufficient to maintain the metastable thermodynamic equilibrium in the timeframe of observation. Therefore, the SCL structure temporarily freezes, producing a noncrystalline and thermodynamically unstable material known as "glass".[1]

The nonequilibrium state of the glass structure is time-dependent; it only seems frozen when observed during a short enough experimental time. Any glass gradually and spontaneously relaxes to a state having lower Gibbs free energy, and eventually reaches the metastable SCL state. The rate at which the structure relaxes depends on the temperature, pressure, chemical composition, and thermobaric history of the glass. Meanwhile, it is possible that other processes such as liquid-liquid phase separation or crystal nucleation and growth occur concurrently during relaxation.

Different types of relaxation processes occur in glasses. The slowest phenomenon is a primary or α-relaxation, whereas the faster β-relaxation refers to secondary relaxation.[2,3] The α-relaxation involves a cooperative rearrangement of the structural units, e.g., Si–O rings in silicate glasses. In contrast, β-relaxation refers to non-collective atomic transport,[4] for instance, self-diffusion of network modifiers, such as alkali and alkaline earth ions. Among these processes, we highlight the structural relaxation in the absence of stress and two types of stress relaxation (externally applied or internal residual stress).

To establish a standard terminology, we will define structural relaxation and stress relaxation following Scherer[5] and Doss et al.[6] Structural relaxation is "*the response of a material subjected to an isothermal hold measured through observable changes in the material's properties […] due to structural rearrangements over time*". Stress relaxation is "*the time-dependent response of the stress developed within a material subjected to a mechanical strain.*" The term structural relaxation commonly appears as volume relaxation, since the relaxation process involves continuous glass volume changes; it has been measured in the literature in classic studies.[7–10]

We want to draw the reader's attention to the difference between "stress" relaxation, as previously defined, and the classical case of "residual stress" relaxation, which is normally relieved by annealing procedures. Residual stresses are generated in glasses due to the different cooling rates to which the surface and interior of an object are subjected, which lead to a density gradient that causes these stresses. They are relieved by an annealing treatment, usually close to the glass transition temperature for 15–20 minutes, followed by slow cooling to avoid additional density gradients. Therefore, residual stress relaxation could be defined as "*the response of a glass subjected to an isothermal hold to reduce residual stress via atomic rearrangements.*" Residual stress relaxation is faster than both stress relaxation and structural relaxation.

Fictive temperature ($T_f$) and fictive pressure are structure-related parameters that can assist with understanding relaxation processes. In this work, the pressure will be kept at ambient; hence, we will deal only with structural changes caused variations in time or temperature. If the structure of a given SCL is in metastable equilibrium at a temperature $T_a$, then its fictive temperature is equal to $T_a$. Suppose this particular material is submitted to a change in temperature, for instance, going from $T_a$ to $T_b$. In that case, its fictive temperature will change to $T_b$ if sufficient time is allowed to the system, reflecting a change in the spatial arrangement of the structural units (α-relaxation).

Knowledge of the relaxation mechanism and kinetics is fundamental in glass science because relaxation (or lack of) is intrinsically related to the vitrification, annealing, and tempering processes and the dimensional stability of glass articles.[11] In addition, it is of vital importance for high technology applications, such as optical fibers,[12] flat panel displays,[13] and chemically strengthened cover glass.[14,15]

The structural relaxation times measured by refractive index, enthalpy, density, and volume thermal expansion for different glasses seem to be similar in magnitude (within the experimental uncertainty) to their shear relaxation time, which is calculated using the Maxwell equation.[16–19] This widespread belief has motivated one of the authors, Zanotto,[20] to estimate the relaxation time at room temperature for stained glass windows in medieval cathedrals, where the extrapolated *equilibrium* shear viscosity data were used to estimate the time required for a $GeO_2$ glass (with $T_g$ similar to that of soda-lime-silica glass) to flow at room temperature. In a follow-up work, Zanotto and Gupta[21] used the *isostructural* shear viscosity to estimate a lower boundary for the relaxation time of a soda-lime-silica glass at room temperature. In related works, a combination of *equilibrium* and *nonequilibrium* shear viscosity was used by Welch et al.[11] and Gulbiten et al.[15] to estimate the relaxation time at room temperature of a Corning Gorilla® glass and a medieval cathedral window glass, respectively. Despite the many orders of magnitude difference observed in these four estimates, their main conclusion was that, at room temperature, the viscosity of these oxide glasses is far too high to explain the thickness variations observed in medieval stained-glass windows. It is worth noting that some authors[6,22–24] have shown that the average structural relaxation times are *longer* than the relaxation times calculated from the Maxwell equation (which relies on shear viscosity) by approximately one order of magnitude for temperatures near and below the $T_g$ range.

In this context, we are interested in finding a more comprehensive answer to the following question: *Is the experimental average structural relaxation time comparable with the average relaxation time calculated from the Maxwell equation?* It is well-known that the classical Tool–Narayanaswamy–Moynihan (TNM) model[25–27] and the modern Mauro–Allan–Potuzak (MAP) model[28] use *combinations* of equilibrium and nonequilibrium shear viscosity to describe glass relaxation kinetics. Here we intend to verify whether one of these two types of viscosity dominates relaxation, or a combination of both is fundamental to describe the process. For this task, we determined the structural relaxation kinetics of a lead metasilicate ($PbSiO_3$, PS) glass by refractive index measurements with the time of isothermal treatment at temperatures below $T_g$. This particular glass composition was chosen because experimental values of both equilibrium and isostructural (hard to measure) viscosity are available, as well as because it is very stable against crystallization in long treatment times.

# 2 Literature review

## 2.1 Summary of the governing equations

One of the simplest equations for the relaxation kinetics is the Kohlrausch equation, also known as stretched exponential function. Historically, Rudolf Kohlrausch in 1854 [29] and his son Friedrich Kohlrausch in 1863 [30] and 1866 [31] were the first to observed that relaxation processes have time dependencies given by Eq. (1). Sometimes this equation is also referred as the Kohlrausch–Williams–Watts (KWW) function, but this terminology is discouraged in this context; it is more appropriate refer to the Fourier transform of the Kohlrausch equation since Williams and Watts transformed this equation analytically.[32,33] The Kohlrausch equation must be used under isobaric and isothermal conditions.

$$\varphi(t, T, T_f) = \frac{p(t, T, T_f) - p_\infty(T)}{p_0(T_f) - p_\infty(T)} = \exp\left[-\left(\frac{t}{\tau_k(T, T_f)}\right)^{\beta(T, T_f)}\right] \quad (1)$$

In Eq. (1), $\varphi$ is the relaxation parameter, which is unity for an unrelaxed system and zero for a relaxed system; $p$ is the value of some property (density, enthalpy, viscosity, or refractive index, for example), which varies during the relaxation process; $p_\infty$ is the equilibrium value of that property at a temperature $T$ of interest; $p_0$ is the value of the property at the initial fictive temperature, $T_f$; $t$ is the experimental time; $\beta$ is the Kohlrausch exponent, which is a positive dimensionless factor, usually below or equal to unity; $\tau_k$ is the characteristic relaxation time.

The experimental average structural relaxation time, $\bar{\tau}_{exp}$, is related to the characteristic relaxation time via Eq. (2). If the liquid follows the Maxwell fluid equation, then its shear viscosity, η, is related to the average shear relaxation time ($\bar{\tau}_\eta$) shown in Eq. (3).

$$\bar{\tau}_{exp}(T, T_f) = \tau_k(T, T_f) \Gamma\left(\frac{1}{\beta(T, T_f)} + 1\right) \quad (2)$$

$$\bar{\tau}_\eta(T) = \frac{\eta(T)}{G_\infty} \quad (3)$$

In the previous equations, Γ is the gamma function and $G_\infty$ is the instantaneous shear modulus, which is practically temperature-independent.[34,35] If the previous equations are assumed to be valid for glasses, we can check for correlations between $\bar{\tau}_{exp}$ and $\bar{\tau}_\eta$, which could lead to important clues on the mechanisms behind structural relaxation and viscous flow.

To model the temperature dependence of the equilibrium shear viscosity, we will use the Vogel–Fulcher–Tammann (VFT),[36–38] Avramov–Milchev (AM),[39] and Mauro–Yue–Ellison–Gupta–Allan (MYEGA)[40] models (Eqs. (4–6), respectively). They are considered here because, among many available equilibrium viscosity models, the VFT and AM models usually provide upper and lower boundaries, respectively, for the viscosity below the laboratory glass transition temperature. At the same time, the MYEGA model seems to offer an accurate prediction of low-temperature viscosity.[40] In this way, we cover an uncertainty window.

$$\log_{10}(\eta(T)) = \log_{10}(\eta_{\infty,VFT}) + \frac{B}{T - T_0} \tag{4}$$

$$\log_{10}(\eta(T)) = \log_{10}(\eta_{\infty,AM}) + \left[\frac{C}{T}\right]^\alpha \tag{5}$$

$$\log_{10}(\eta(T)) = \log_{10}(\eta_{\infty,MYEGA}) + [12 - \log_{10}(\eta_{\infty,MYEGA})]\frac{T_g}{T}\exp\left[\left(\frac{m}{12-\log_{10}(\eta_{\infty,MYEGA})} - 1\right)\left(\frac{T_g}{T} - 1\right)\right] \tag{6}$$

In Eqs. (4–6), $\eta_{\infty,VFT}$, $B$, $T_0$, $\eta_{\infty,AM}$, $C$, $\alpha$, $\eta_{\infty,MYEGA}$, $T_g$, and $m$ are adjustable parameters that depend on the chemical composition of the glass. It is relevant to note that $T_0$ is the predicted temperature in the VFT framework at which the equilibrium shear viscosity diverges to infinity. Thus, within this framework, the glass is not able to relax at temperatures equal to or below $T_0$. However, $T_0$ is typically well below the range of measurable relaxation times, and will not play a significant role in the present analysis.

## 2.2 Shear versus structural relaxation

Several relaxation processes could be controlled by diffusion processes that are akin to volume viscous flow rather than to shear flow. Unfortunately, only a handful of data are available for volume (bulk) viscosity.[23] It is not an easy task to compare structural and shear relaxation times because of the measurement uncertainty. In fact, the Maxwell equation has been widely used in the literature, and the experimental average structural relaxation time is often considered to coincide with the (calculated) average shear relaxation time.[41] Therefore, small differences between these relaxation times are commonly attributed to the measurement uncertainty.[17,19]

In 1976, Moynihan et al.[24] provided a dataset comparing structural and shear relaxation using the viscous lubricant 5-phenyl-4-ether (5P4E, $C_6H_5(OC_6H_4)_3OC_6H_5$). The average structural relaxation times were calculated using Eq. (2), with $\tau_k$ obtained from the Kohlrausch equation by measurements of the digital correlation spectroscopy as a function of time. The average shear relaxation times were calculated through the Maxwell equation using equilibrium shear viscosity and $G_\infty$ (Eq. (3)). Figure 1a shows the results obtained by Moynihan. Structural relaxation takes approximately 1.0 to 2.0 orders of magnitude longer than shear relaxation in a range of temperatures above $T_g$, indicating that volume viscosity is typically larger than shear viscosity. A similar result was reported by Simmons et al.[22] for an oxide glass former with composition $70.5SiO_2 \cdot 22.7B_2O_3 \cdot 6.8Na_2O$.

Doss et al.[6] recently compared structural and stress relaxation times at temperatures over 100 K below $T_g$ for a Corning Jade® glass. The relaxation times were also calculated via the Maxwell equation using equilibrium shear viscosity and $G_\infty$. The experimental average structural relaxation times were obtained by measuring the variation of density as a function of time. The average stress relaxation times were measured by subjecting fully equilibrated samples at the measurement temperature to a three-point bend with a constant strain. Their results are summarized in Figure 1b. It is clear that structural relaxation is *much* *longer* than nonequilibrium (MAP model) shear relaxation, and takes approximately 1.0 to 1.5 orders of magnitude *longer* than equilibrium (MYEGA model) shear relaxation. Furthermore, they showed that the relaxation times, calculated by the Maxwell equation, correspond to a stress relaxation time.

Moynihan et al.[24] compared the average relaxation times inferred from different properties in the glass transition region. Some of their data are shown in Table 1 for two glasses: $B_2O_3$ and $As_2Se_3$. In general, the structural relaxation time is *longer* than the shear relaxation time by a factor of 2–15 in the glass transition region. Additionally, Moynihan's data show that β depends on the property being analyzed.

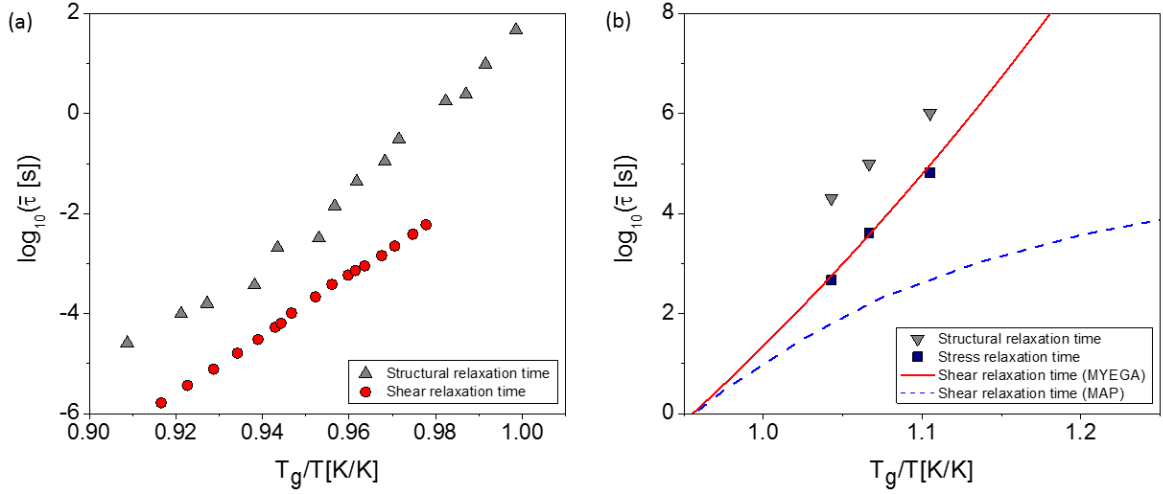

Figure 1 (a) Temperature dependence of the average structural and shear relaxation times for 5P4E ($T_g$ = 250 K). (b) Corning Jade® glass ($T_g$ = 1075 K), comparison between the structural, shear, and stress relaxation kinetics. Data adapted from the literature[6,24]

Table 1 Comparison of average relaxation times and the Kohlrausch exponent for different properties in the glass transition region for two glasses[24]

| Glass | Relaxing property | In response to changes in | T [K] | $\overline{\tau}(T)$ [s] | β(T) |
|---|---|---|---|---|---|
| $B_2O_3$ | Refractive index | Temperature | 536 | 17000 | 0.83 |
| | Enthalpy | Temperature | 536 | 9700 | 0.65 |
| | Volume | Pressure | 536 | 6700 | 0.60 |
| | Shear stress | Shear strain | 536 | 1100 | – |
| $As_2Se_3$ | Enthalpy | Temperature | 450 | 540 | 0.67 |
| | Volume | Temperature | 450 | 100 | 0.80 |
| | Shear stress | Shear strain | 450 | 45 | – |

## 2.3 Equilibrium and isostructural viscosity

The relaxation phenomenon is linked to changes in the distribution of the structures in glass. If the structure does not change within the time frame of the experiment, then viscous processes are governed by the so-called isostructural viscosity (ISV). However, the isostructural regime is limited to the very early stages of relaxation, an extreme situation that can only be measured when structural rearrangements are sufficiently slow. After the structure begins to change, and before the equilibrium structure is reached, the viscous processes are governed by the so-called nonequilibrium viscosity.[28] As the structure changes during relaxation, so does the nonequilibrium contribution, as shown schematically in Figure 2 by an arrow—the frozen glass structure changes with time towards the metastable equilibrium SCL state.

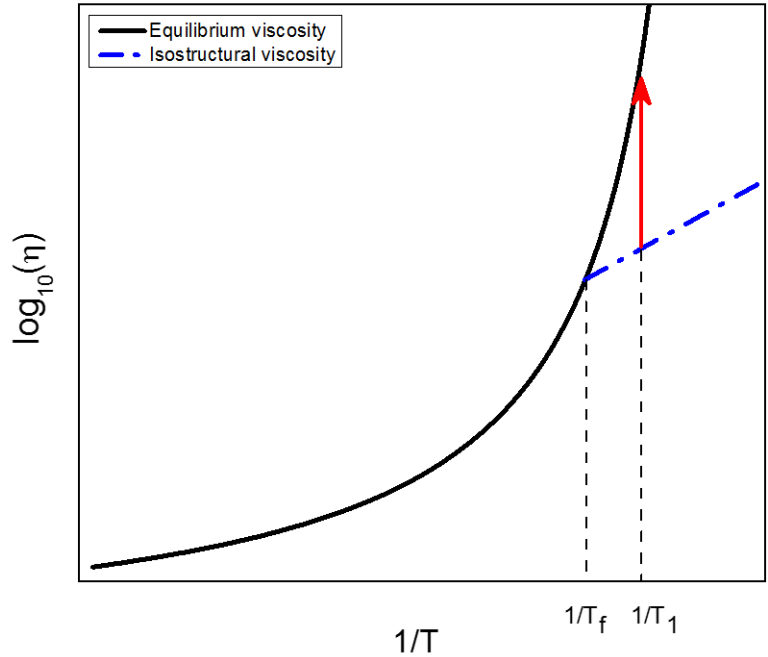

Figure 2 Schematic viscosity *vs*. inverse temperature plot. The solid line is the equilibrium shear viscosity. The dash-dot line represents the ISV with a fictive temperature equal to $T_f$. The arrow indicates the relaxation process, i.e., the change of viscosity with time toward equilibrium in a temperature $T_1$

Gupta and Heuer[42] provided a thorough review of ISV, describing the phenomenon, relevant experimental results available in the scarce literature, and providing a useful discussion on ISV's existing phenomenological models and their respective limitations. According to those authors, "*isostructural viscosity is the viscosity at a temperature T of a glass having the structure of the liquid in equilibrium at a different temperature $T_f$.*"

In their work, Gupta and Heuer critically analyzed the literature data on ISV contemplating various systems, such as oxides, metals, polymers, and a molecular liquid. From the observation of available experimental data, they listed four key features of isostructural viscosity:

i. ISV is nearly Arrhenius;

ii. At the fictive temperature, the isostructural activation enthalpy is less than the equilibrium activation enthalpy;

iii. The isostructural activation enthalpy decreases with increasing the fictive temperature;

iv. The pre-exponential factor for the equilibrium shear viscosity, $\eta_\infty$, is greater than or equal to the pre-exponential factor for the ISV, $\eta_{\infty iso}$, of the same glass.

Equilibrium shear viscosity has been widely studied by many authors using different techniques. Figure 3 shows the PS viscosity measured by different authors[43–47] and previously reported by us for the same PS batch of this work.[48] Koide et al.[49] measured the ISV for PS glass far below $T_g$ using the fiber-bending method. In this method, the glass fiber is bent in isothermal heat treatment. Then the load is removed and the remaining curvature radius is measured. To calculate the ISV, they used the fiber's radius, radius of curvature of the bending fiber, radius of the deformed fiber after treatment, the stress and strain calculated from the remaining curvature radius, and the Young modulus. The early stages of relaxation were measured by following the viscosity as a function of treatment time. The ISV data shown in Figure 3 correspond to the longer treatment time analyzed by them, i.e., 24 h of treatment time, which is well below the time needed to reach the equilibrium value since the analyses were performed approximately 200 K below $T_g$.

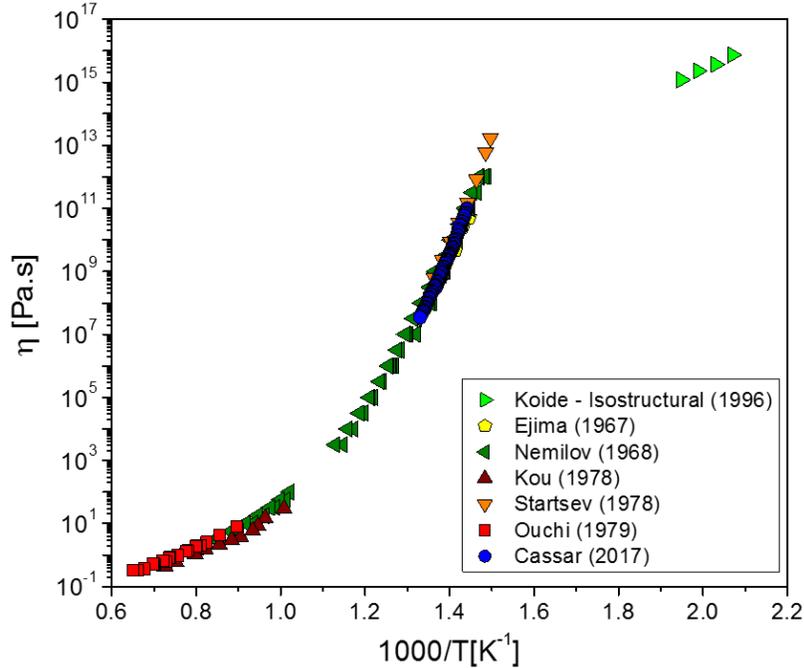

Figure 3 Isostructural and equilibrium shear viscosity of lead metasilicate. Data obtained by different authors using different techniques and glass batches[43–49]

## 3  Materials and Methods

We used samples of a PS glass previously obtained in our laboratory. The melting and casting procedure was described in reference,[48] and other studies on this particular batch of PS glass have been reported in references.[50,51] The glass transition temperature was measured by the onset of the glass–SCL transition of the differential scanning calorimetry (DSC, Netzsch 404 equipment) at a heating rate of 10 K.min$^{-1}$, $T_g^{DSC}$ = 681 K. This particular composition was chosen because experimental values of isostructural viscosity[49] are available.

## 3.1 Viscosity measurements

Equilibrium shear viscosity was measured in a temperature range close to the glass transition. The apparatus used was a homemade penetration viscometer.[52] The method consists in obtaining the isothermal rate of penetration of a rigid indenter into a sample under a known load. The measurements were performed using a sample of approximately 10×10×3 mm. The shear viscosity was computed by using

$$\eta = m \frac{[1-\mu]P}{2\pi^{1/2}R\nu}, \tag{7}$$

where $\mu$ is the Poisson's ratio, $P$ is the applied load, $R$ is the indenter base radius, $m$ is a constant equal to $16/(3\pi^{3/2})$ for cylindrical indenters, and $\nu$ is the penetration rate.[53]

We used a cylindrical indenter with $R = 0.493(1)$ mm and a combination of loads from 10 to 210 N to allow measurement of viscosity above and below the glass transition region. The Poisson's ratio used in the calculation was 0.5, because we measured the SCL's equilibrium shear viscosity. In these measurements, only the data for which the penetration rate was constant were used, which implies a relaxed liquid.

The metallic indenter was exposed to reasonably high stress, but no noticeable creep was observed after the experiments. A high precision linear variable differential transformer was used to reach a penetration resolution of approximately 1 μm. By tailoring these conditions, in the lowest temperature tested, it was possible to get a penetration rate of 774(5) nm/h.

## 3.2 Refractive index measurements

We studied the structural relaxation kinetics by measuring changes in the refractive index with respect to the time of treatment in isothermal experiments. The sample had approximately 10×10×2 mm with two perpendicularly polished faces. The refractive index measurements were made at room temperature in five replicates using a Carl Zeiss Jena Pulfrich-refractometer PR2 with a spectral mercury lamp and a monochromatic green e-line of $\lambda = 546.1$ nm. We used a VeF4 prism with a refractive index of 1.93493(1) (e-line), measuring indices in the range of 1.47–2.01.

Determination of the refractive index was based on the deviation angle measurement, as shown schematically in Figure 4. The sample was placed into the prism with immersion oil to ensure good optical contact between the sample and the prism. The refractive index $n_\lambda$ of the sample is given by

$$n_\lambda^2 = N_\lambda^2 - \cos(\gamma)\left[N_\lambda^2 - \cos^2(\gamma)\right]^{\frac{1}{2}}, \tag{8}$$

where $N_\lambda$ is the refractive index of the prism at the wavelength of illumination and $\gamma$ is the angle of the refracted beam.

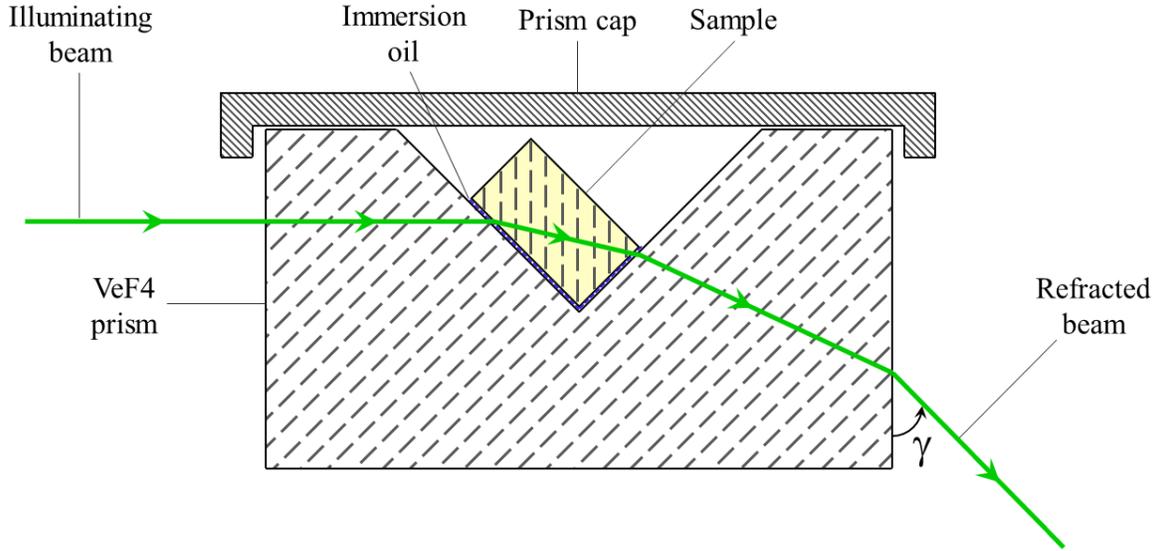

Figure 4 Schematic representation of the refractometer with a sample

## 3.3 Structural relaxation kinetics

The structural relaxation kinetics were followed as a function of time at 5, 10, 15, 20, and 25 K below $T_f$. The first step in the experimental procedure consisted in holding the sample at 5 K below its $T_g^{DSC}$ for 4 h, which was sufficient to relax the structure to its metastable equilibrium (SCL) configuration, fixing the initial $T_f$. After removing the sample from the furnace, a quick temperature drop froze the fictive temperature at $T_f = 676$ K.

The measurements were carried out by alternating isothermal treatments and room temperature refractive index determinations over time until the refractive index value was stabilized. Prior to all refractive index measurements, the sample was stabilized at the equipment's room temperature for 15 min.

The experiments were performed using the same sample for each treatment temperature. After all the necessary experiments were conducted at a given temperature, the initial $T_f$ had to be "reset" by redoing the heat treatment at 676 K for 4 h. Thus, the starting refractive index value (for $t = 0$ s) is the same for all experiments, i.e., at the equilibrium value of the initial $T_f$.

The parameters $\beta$, $\tau_k$, $n_0$, and $n_\infty$ were obtained by performing nonlinear regressions of Eq. (1) on the experimental data, which was done separately for each isothermal dataset. With $\tau_k$ and $\beta$, the $\bar{\tau}_{exp}$ were computed at each temperature by Eq. (2). The $\bar{\tau}_\eta$ were computed by Eq. (3) using the shear viscosity data and a value of $G_\infty = 17.51$ GPa.[54]

The confidence intervals of the Kohlrausch equation were calculated considering statistical confidence of 2σ (yielding a probability of about 95% of the true value within this range).

# 4 Results and discussion

Figure 5a shows the experimental shear viscosity data measured in this work and from the literature together with AM, VFT, and MYEGA regressions. The data measured above and below $T_g$ are in good agreement with the literature. By employing a high load and measuring the indenter displacement for days below $T_g$, we could expand the existing experimental viscosity curve in one order of magnitude, up to $10^{14}$ Pa.s.

We extrapolated the ISV data measured by Koide et al., assuming that they follow an Arrhenius equation, as Gupta and Heuer have suggested. The extrapolation values seem to contradict the feature *iv* observed by Gupta and Heuer, since the $\log_{10}(\eta_{\infty iso} [\text{Pa.s}]) = 2.7(7)$ is greater than $\log_{10}(\eta_{\infty,AM} [\text{Pa.s}]) = -0.83(5)$, $\log_{10}(\eta_{\infty,VFT} [\text{Pa.s}]) = -3.50(7)$, and $\log_{10}(\eta_{\infty,MYEGA} [\text{Pa.s}]) = -1.63(6)$. However, special care is required when dealing with such massive extrapolation because of the considerable uncertainty of this procedure, as can be visually observed in Figure 5b.

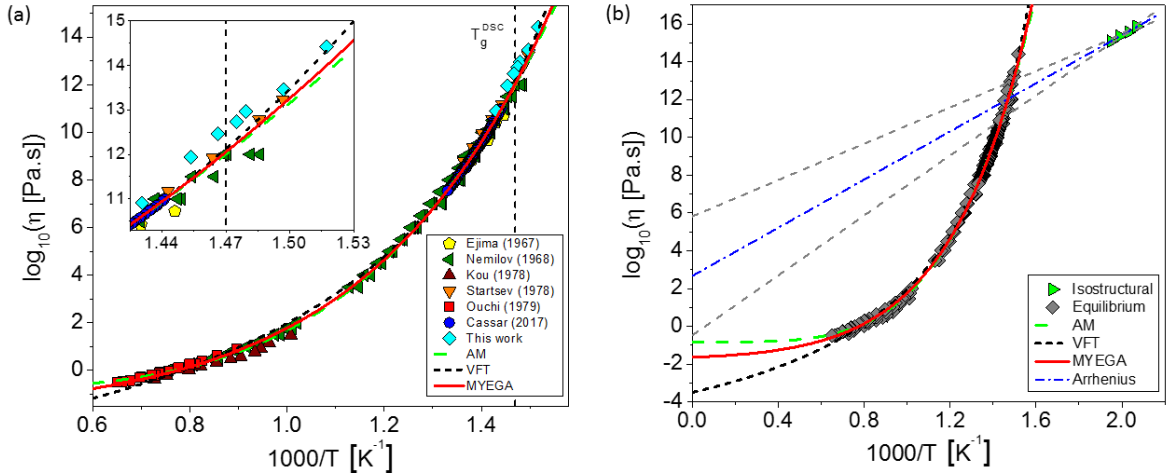

Figure 5 (a) Equilibrium shear viscosity data measured in this work (cyan diamonds) combined with literature data[43–48] for lead metasilicate. The AM, VFT, and MYEGA regressions were obtained using all equilibrium shear viscosity data. The insert is a zoom in the temperature range where we measured the viscosity. The uncertainty bars of the data measured in this work are smaller than the symbol size. (b) Isostructural data[49] with an Arrhenius extrapolation. The gray dashed lines show the 95% confidence interval. Values measured in this study are provided in the Supplemental Material, Table S1

Figure 6 shows the structural relaxation kinetics at five temperatures below the glass transition temperature, observed by the change in the refractive index, which was precise enough to capture the subtle structural changes during relaxation. The time required for the glass structure to fully relax at 5 K below $T_f$ is in the order of one day. However, only 20 K below this temperature, the time required for the same to occur is about one month! These observations put the exponential nature of structural relaxation into perspective and show that measurements below the reported temperature range for PS glass are challenging and may lead to prohibitively long treatment times.

Refractive index is a temperature-dependent property; thus, its value at the temperature of any of the isothermal treatments is different from its room temperature value.[55,56] However, the relevant result here is how these values systematically change with the different heat treatments, which gives us an indirect way to probe the structural changes in the glass. Moreover, Figure 6 clearly shows different values of room temperature refractive index for a single sample. This observation is essential to clarify that the value of the analyzed property in the glassy state depends on the thermal history of the sample.

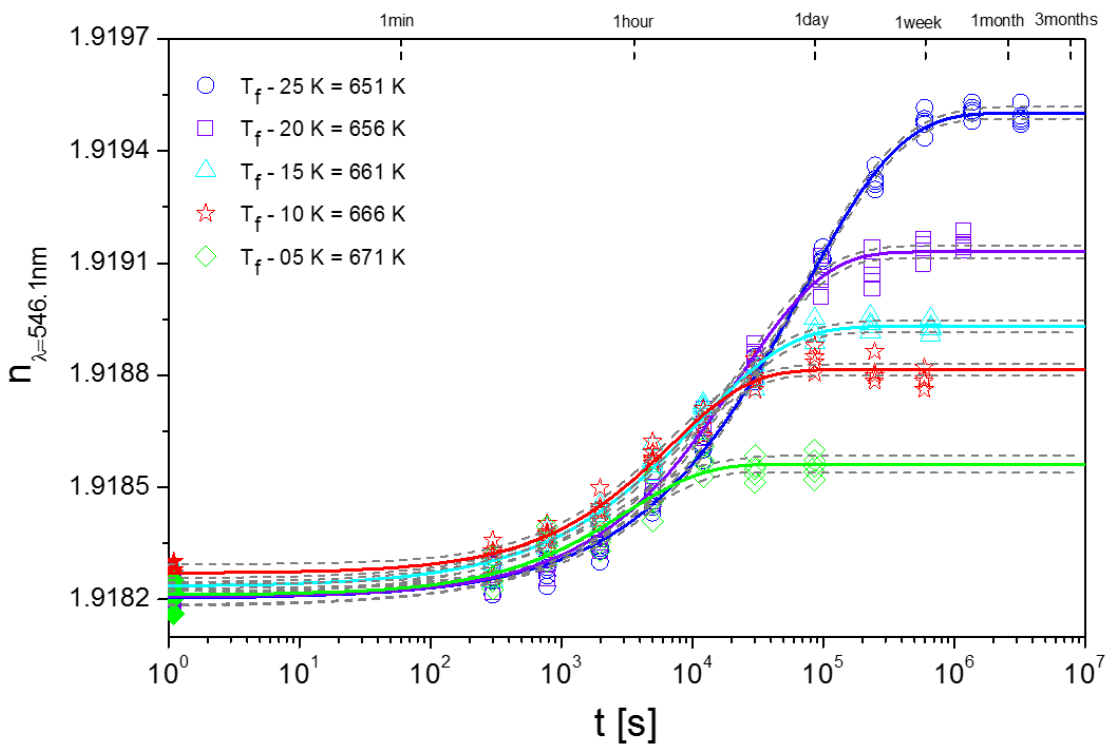

Figure 6 Room temperature measurements of the refractive index of PS glass as a function of treatment time. Five different isothermal heat treatments below the fictive temperature are shown. The filled symbols represent the values of the refractive indices measured at $t = 0$ s. The solid lines are the Kohlrausch equation regressions, and the dashed lines show the 95% confidence interval. Values are provided in the Supplemental Material, Table S2

Figure 6 shows that the Kohlrausch equation describes well the relaxation kinetics. Figure 7 shows the adjustable parameters obtained from the regressions. Linear regressions were performed with the parameters using a 95% confidence level. The average refractive index at the initial fictive temperature ($n_0$) is 1.91822(3), where the uncertainty is the calculated standard deviation from data points. The other parameters are significantly influenced by temperature. The Kohlrausch exponent ($\beta$) is related to the distribution of relaxation times of atomic groups in the material. We observe that $\beta$ increases with temperature, corroborating the literature reports for other glasses.[57] The equilibrium values of the refractive index ($n_\infty$) decrease with temperature, which is in agreement with several studies, including the classic work by Spinner and Napolitano.[58] Finally, $\tau_k$ exponentially decreases with temperature, as expected for the relaxation process.

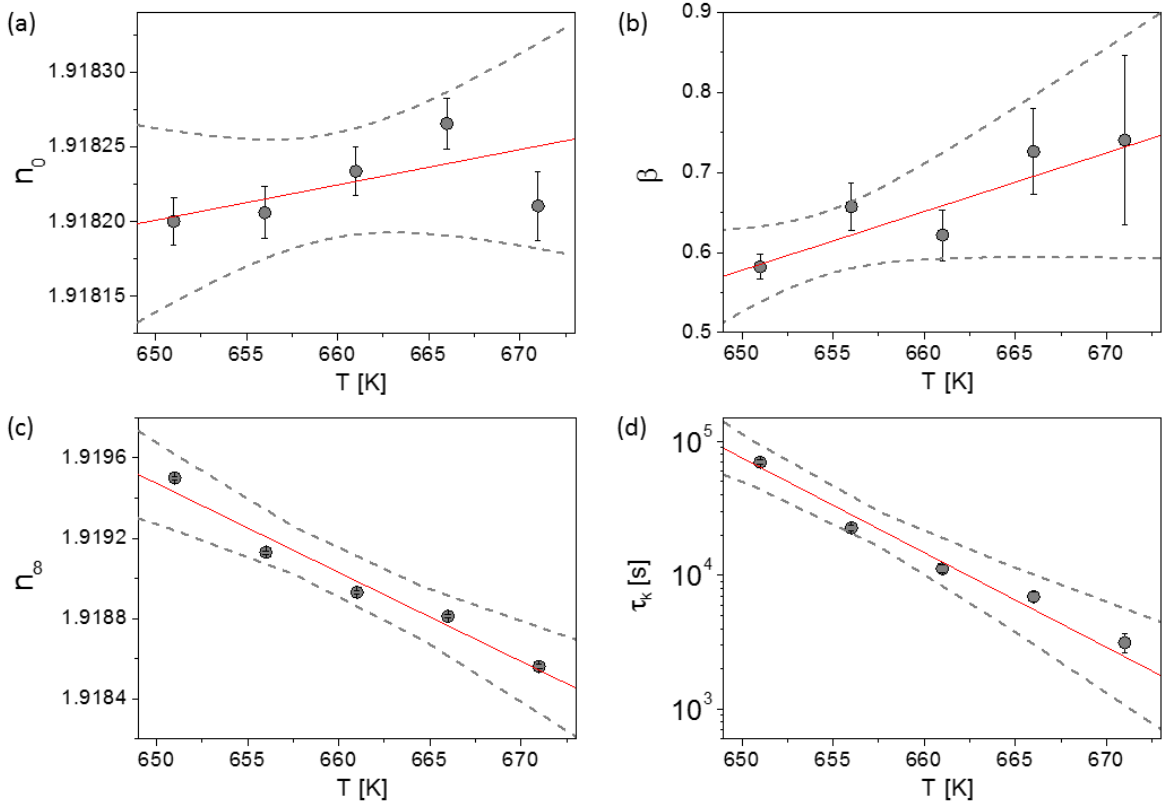

Figure 7 Parameters related to the relaxation process plotted against the temperature of the isothermal heat-treatments. (a) Refractive index after "resetting" the structure for 4h at T = 676 K (no correlation with the heat treatment temperature is expected), (b) Kohlrausch exponent, (c) equilibrium refractive index at temperature T, and (d) characteristic relaxation time. The gray dashed lines show the 95% confidence interval

Figure 8a shows the average relaxation times calculated from equilibrium shear viscosity, isostructural viscosity, and refractive index measurements. As expected, the ISV has a temperature dependence widely different from that of the equilibrium shear viscosity. The temperature dependence and magnitude of the average structural relaxation times (as inferred from the refractive index measurements) are thus much closer to those of the equilibrium shear viscosity than to those of the ISV, but not precisely equal. This analysis corroborates those of reference[6] for Jade® glass at much deeper undercoolings.

A closer detailed look at Figure 8b evidences that the experimental relaxation times lie *above* the (calculated) relaxation kinetics via equilibrium shear viscosity. These results suggest that structural relaxation might be described by volume viscosity and bulk modulus instead of by shear viscosity and shear modulus, which is in agreement with the conclusions of Doss et al.[6] This result is in accordance with references[6,22,24] and the results of Rekhson,[23] who reported that structural relaxation time takes longer than shear relaxation time by a factor of 4–20 in the glass transition region. In conclusion, shear viscosity and shear modulus can only give a lower boundary for the average structural relaxation times when used with the Maxwell equation.

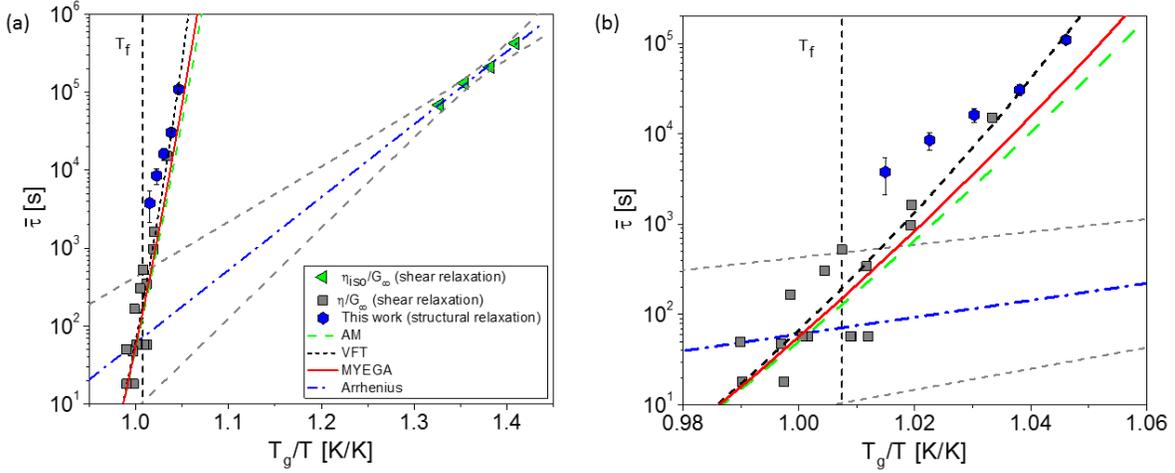

Figure 8 Comparison between $\bar{\tau}_{exp}$ (structural relaxation) and $\bar{\tau}_\eta$ (shear relaxation). $T_f$ indicates the fictive temperature initially set at 676 K. The gray dashed lines show the 95% confidence interval. (a) Overall picture showing the equilibrium and isostructural data. (b) Zoom at temperatures near $T_f$

A final remark is that the temperature dependence of the experimental average structural relaxation times and average shear (Maxwell equation) relaxation times seems somewhat *different*.

## 5 Conclusions

We have investigated the kinetics of structural relaxation of a lead metasilicate glass by measuring the isothermal changes of the refractive index over time. The Kohlrausch equation describes the experimental structural relaxation kinetics at temperatures between 5 and 25 K below the respective $T_f$ (which was initially set 5 K below the glass transition temperature). The time required for the refractive index to reach equilibrium (99.9% relaxation) varied from a few hours to 20 days in this temperature range. The Kohlrausch exponent (a fitting parameter) shows a positive correlation with temperature, as expected from theoretical grounds and literature results for other glasses.

We have expanded the data range of experimental viscosity by one order of magnitude, reaching $10^{14}$ Pa.s, and our new results are in good agreement with previously reported data. The experimental average structural relaxation times obtained via refractive index measurements are somewhat *longer* than the values calculated from equilibrium shear viscosity via the Maxwell equation. Both structural and shear relaxation times are *much longer* than the relaxation times computed using isostructural viscosity. The temperature dependence of these relaxation times seems to be somewhat *different*.

Hence, the frequently used Maxwell equation, which relies on equilibrium shear viscosity and shear modulus, underestimates the average structural relaxation time. These novel results for lead metasilicate glass corroborate recent findings for Jade® glass. They suggest that structural relaxation might be described by volume viscosity and bulk modulus rather than by shear viscosity and shear modulus.

## Acknowledgments

The authors are grateful to the constructive insights by Prabhat K. Gupta, John C. Mauro, Collin J. Wilkinson, Marcos Jardel Henriques, Oilson Alberto Gonzatto Junior, and Francisco Louzada Neto. This research was performed using funding received from the São Paulo Research Foundation, FAPESP, Grant Nos. 2013/07793-6 (CEPID), 2019/15108-8 (RFL), and 2017/12491-0 (DRC) and the Brazilian National Research Council, CNPq, Grant No. 130495/2019-0 (RFL). This study was financed in part by the Coordenação de Aperfeiçoamento de Pessoal de Nível Superior - Brasil (CAPES) - Finance Code 001.

# Supplemental Material to "Is the Structural Relaxation of Glasses Controlled by Equilibrium Shear Viscosity?"

Ricardo Felipe Lancelotti[1,2,*], Daniel Roberto Cassar[2], Marcelo Nalin[3], Oscar Peitl[2], Edgar Dutra Zanotto[2,*]

[1]Federal University of São Carlos, Graduate Program in Materials Science and Engineering, 13565-905, São Carlos, SP, Brazil.
[2]Center for Research, Technology and Education in Vitreous Materials, Department of Materials Engineering, Federal University of São Carlos, 13565-905, São Carlos, SP, Brazil.
[3]Institute of Chemistry, São Paulo State University, UNESP, 14800-060, Araraquara, SP, Brazil.

* lancelotti.r@dema.ufscar.br, dedz@ufscar.br

## Supplemental Material

Tables S1, and S2 show all the experimental data obtained from this study.

Table S1: Experimental viscosity data obtained from this study.

| T [K] | $\eta$ [Pa.s] |
|---|---|
| 659 | $2.65(1)\times10^{14}$ |
| 668 | $2.86(1)\times10^{13}$ |
| 676 | $9.2(1)\times10^{12}$ |
| 678 | $5.34(3)\times10^{12}$ |
| 682 | $2.90(2)\times10^{12}$ |
| 688 | $8.76(2)\times10^{11}$ |
| 699 | $8.04(2)\times10^{10}$ |

Table S2: Experimental room temperature refractive index data as a function of treatment time obtained from this study.

| T [K] | t [s] | $n_{\lambda=546.1nm}$ | | | | |
|---|---|---|---|---|---|---|
| 671 | 0 | 1.91821 | 1.91821 | 1.91825 | 1.91816 | 1.91822 |
| 671 | 300 | 1.91825 | 1.91826 | 1.91832 | 1.91823 | 1.91826 |
| 671 | 780 | 1.91834 | 1.91835 | 1.91840 | 1.91830 | 1.91834 |
| 671 | 1980 | 1.91835 | 1.91837 | 1.91841 | 1.91833 | 1.91838 |
| 671 | 4980 | 1.91846 | 1.91847 | 1.91853 | 1.91841 | 1.91846 |
| 671 | 12180 | 1.91856 | 1.91857 | 1.91861 | 1.91853 | 1.91857 |
| 671 | 30180 | 1.91856 | 1.91855 | 1.91859 | 1.91851 | 1.91853 |
| 671 | 85980 | 1.91856 | 1.91857 | 1.91860 | 1.91852 | 1.91854 |
| 666 | 0 | 1.91827 | 1.91828 | 1.91830 | 1.91824 | 1.91827 |
| 666 | 300 | 1.91833 | 1.91832 | 1.91836 | 1.91826 | 1.91832 |
| 666 | 780 | 1.91838 | 1.91837 | 1.91840 | 1.91833 | 1.91837 |
| 666 | 1980 | 1.91845 | 1.91843 | 1.91850 | 1.91841 | 1.91844 |
| 666 | 4980 | 1.91859 | 1.91858 | 1.91862 | 1.91856 | 1.91857 |
| 666 | 12180 | 1.91867 | 1.91867 | 1.91871 | 1.91864 | 1.91867 |
| 666 | 30180 | 1.91880 | 1.91878 | 1.91884 | 1.91876 | 1.91879 |
| 666 | 86380 | 1.91883 | 1.91885 | 1.91888 | 1.91881 | 1.91883 |
| 666 | 248380 | 1.91880 | 1.91881 | 1.91886 | 1.91878 | 1.91881 |
| 666 | 593980 | 1.91879 | 1.91878 | 1.91882 | 1.91876 | 1.91878 |
| 661 | 0 | 1.91826 | 1.91829 | 1.91822 | 1.91824 | 1.91824 |
| 661 | 300 | 1.91829 | 1.91828 | 1.91831 | 1.91825 | 1.91828 |
| 661 | 780 | 1.91836 | 1.91837 | 1.91839 | 1.91831 | 1.91834 |
| 661 | 1980 | 1.91842 | 1.91845 | 1.91841 | 1.91843 | 1.91844 |
| 661 | 4980 | 1.91856 | 1.91856 | 1.91862 | 1.91853 | 1.91857 |
| 661 | 12180 | 1.91872 | 1.91870 | 1.91872 | 1.91864 | 1.91871 |
| 661 | 30180 | 1.91881 | 1.91879 | 1.91883 | 1.91876 | 1.91879 |
| 661 | 86380 | 1.91891 | 1.91892 | 1.91895 | 1.91889 | 1.91892 |
| 661 | 230380 | 1.91894 | 1.91893 | 1.91896 | 1.91892 | 1.91894 |
| 661 | 662380 | 1.91893 | 1.91892 | 1.91895 | 1.91891 | 1.91892 |
| 656 | 0 | 1.91821 | 1.91822 | 1.91826 | 1.91818 | 1.91821 |
| 656 | 300 | 1.91826 | 1.91825 | 1.91829 | 1.91822 | 1.91825 |
| 656 | 780 | 1.91831 | 1.91829 | 1.91834 | 1.91826 | 1.91831 |
| 656 | 1980 | 1.91837 | 1.91836 | 1.91839 | 1.91833 | 1.91835 |
| 656 | 4980 | 1.91850 | 1.91849 | 1.91854 | 1.91846 | 1.91850 |
| 656 | 12180 | 1.91866 | 1.91867 | 1.91871 | 1.91863 | 1.91866 |
| 656 | 30180 | 1.91886 | 1.91884 | 1.91889 | 1.91882 | 1.91886 |
| 656 | 96240 | 1.91905 | 1.91906 | 1.91912 | 1.91901 | 1.91906 |
| 656 | 236640 | 1.91909 | 1.91907 | 1.91914 | 1.91903 | 1.91909 |
| 656 | 582240 | 1.91913 | 1.91915 | 1.91916 | 1.91910 | 1.91915 |
| 656 | 1187040 | 1.91915 | 1.91916 | 1.91919 | 1.91913 | 1.91914 |
| 651 | 0 | 1.91822 | 1.91823 | 1.91825 | 1.91821 | 1.91823 |
| 651 | 300 | 1.91825 | 1.91826 | 1.91829 | 1.91821 | 1.91825 |
| 651 | 780 | 1.91826 | 1.91828 | 1.91831 | 1.91823 | 1.91829 |
| 651 | 1980 | 1.91834 | 1.91834 | 1.91837 | 1.91830 | 1.91833 |
| 651 | 4980 | 1.91845 | 1.91845 | 1.91848 | 1.91843 | 1.91845 |
| 651 | 12180 | 1.91862 | 1.91860 | 1.91864 | 1.91857 | 1.91861 |
| 651 | 30180 | 1.91882 | 1.91883 | 1.91885 | 1.91878 | 1.91883 |
| 651 | 99480 | 1.91911 | 1.91911 | 1.91914 | 1.91909 | 1.91911 |
| 651 | 250680 | 1.91933 | 1.91931 | 1.91936 | 1.91930 | 1.91932 |
| 651 | 596280 | 1.91947 | 1.91949 | 1.91952 | 1.91943 | 1.91948 |
| 651 | 1373880 | 1.91952 | 1.91950 | 1.91953 | 1.91948 | 1.91951 |
| 651 | 3238680 | 1.91948 | 1.91949 | 1.91953 | 1.91947 | 1.91949 |